\documentclass[12pt]{article}
\usepackage{latexsym}
\usepackage{graphicx}

\catcode`\@=11
\@addtoreset{equation}{section}

\catcode`@=12
\relax

\newcommand{\nn}{\nonumber}

\begin{document}
\thispagestyle{empty}

\renewcommand{\thefootnote}{\fnsymbol{footnote}}
\font\csc=cmcsc10 scaled\magstep1
{\baselineskip=14pt
 \rightline{
 \vbox{\hbox{TIT-HEP-488}
       \hbox{November 2002}
       \hbox{hep-th/0211287}
}}}

\vfill
\begin{center}
{\LARGE\bf
 $\mathbf{\mathcal{N}=1}$ Gauge Theory with Flavor}
\\
\vspace{3mm}
{\LARGE\bf   from Fluxes
}
\vspace{5mm}
\\

\vfill

\textsc{ Yutaka Ookouchi}\footnote{
      e-mail address : ookouchi@th.phys.titech.ac.jp}\\
\vskip.2in

{\large \baselineskip=15pt
\vskip.1in
  {\it Department of Physics, 
  Tokyo Institute of Technology,\\
  Tokyo 152-8511, Japan}
\vskip.1in
}

\end{center}
\vfill

\begin{abstract}
{


Cachazo and Vafa studied $\mathcal{N}=1$ dynamics of $U(N)$ gauge theory from a viewpoint of type IIB superstring compactified on a Calabi-Yau manifold with fluxes. They proved the equivalence between the dynamics and that of $\mathcal{N}=2$ supersymmetric $U(N)$ gauge theory deformed by certain superpotential terms. We generalize their results to gauge theories with massive flavors in fundamental representation for classical gauge groups. When the additional tree level superpotential takes the form of the square of an adjoint chiral superfield we derive Affleck-Dine-Seiberg potentials. By turning off the flux, we obtain the Seiberg-Witten curves of $\mathcal{N}=2$ gauge theories. 

}
\end{abstract}
\vfill

\setcounter{footnote}{0}
\renewcommand{\thefootnote}{\arabic{footnote}}
\newpage
\setcounter{page}{1}


\section{Introduction}

Recently, Dijkgraaf and Vafa proposed that holomorphic information in 
$\mathcal{N}=1$ gauge 
theories with classical gauge groups are derived from matrix models \cite{DV}.
 They claimed that the nonperturbative aspect of holomorphic information was captured by matrix model perturbation. Generalization of DV duality to other gauge theories 
with massive flavors was discussed in \cite{Sweden,McGreevy,Suzuki,BR}. We will derive the results discussed in these papers from a geometric point of view. Dijkgraaf and Vafa have reached this duality via a string theory route using a model with one adjoint matter. For that reason, we specifically address the generalization of string duality to a model with massive flavors.

In \cite{Vafa,CIV} a large $N$ dual description of $\mathcal{N}=2$ $U(N)$  gauge theory deformed
 by certain tree level superpotential was realized using type IIB string theory compactified on Calabi-Yau threefold with fluxes.  The related subject for a type IIB superstring has been discussed in \cite{CKV,CFIKV,EOT,FO,Oh-Tatar}.
In the resolved geometry, world-volume theory of $D5$ branes wrap on $\mathbf{S^2}$'s leads to $\mathcal{N}=1$ supersymmetric gauge theory. After conifold transition, 
$\mathbf{S^2}$ shrinks
 and $\mathbf{S^3}$ appears in the dual geometry. On this dual geometry, 
there are $3$-form fluxes 
through $\mathbf{S^3}$'s that arise from the $D$ brane charge. As discussed 
in \cite{Gukov,TV}
 these fluxes generate the effective superpotential written by glueball superfields. 
Low energy quantities, such as expectation values of adjoint chiral superfield and coupling constant, are given by 
extremization of the effective superpotential. On the 
other hand, purely from the viewpoint of field theory, it is given by specializing to the 
appropriate factorization 
locus of the Seiberg-Witten curve. In \cite{CV}, Cachazo and Vafa proved the equivalence of two 
descriptions for a model with one adjoint chiral superfield.

This paper generalizes Cachazo and Vafa's results to the gauge theory with $N_f$ massive flavors in fundamental representation for classical gauge groups.
This generalization shows a Riemann surface, namely a degenerated Seiberg-Witten curve, that has the same 
genus as a no flavor case but different flux. Turning on various fluxes on the same Riemann surface, we can realize gauge theories with various massive flavors. For $SO/Sp$ theories, because Riemann surface corresponding to two theories are the same, we can realize these two gauge theories with various flavors in terms of fluxes on the same surface. On the other hand, turning off the fluxes in terms of a certain limit, we obtain the Seiberg-Witten curves from Calabi-Yau geometry with fluxes.

This paper is organized as follows: In section 2, we discuss geometric engineering for gauge theories with massive flavors in fundamental representation of gauge groups. Then we see the geometric transition for these models and the effect of flux in the dual geometry. In section 3, we first  discuss the effective superpotential from a purely field theory viewpoint. Next we discuss effective superpotential which is generated from fluxes on a Calabi-Yau manifold. Comparing these two analyses, we see the equivalence of the two results. We give one-form flux explicitly on the Riemann surface in section 4. In section 5, we derive the Affleck-Dine-Seiberg potential from Calabi-Yau manifold with flux. In section 6, turning off the flux, we reproduce Seiberg-Witten curves for the $\mathcal{N}=2$ gauge theories.

\noindent
Note added:  After completion of this note, we received \cite{NSW}, containing the Seiberg-Witten curve derived from the matrix model context. It also indicated the relation to the Calabi-Yau manifold with flux.

\section{Geometric Transition and Dual Description}
\subsection{Geometric engineering}
As discussed in \cite{CIV}, $4$ dimensional $U(N)$ gauge theories with massive chiral multiplets in the fundamental representation are obtained in terms of geometric engineering. First, we review the simplest case, in which the tree level superpotential for an adjoint chiral superfield $\Phi$ is given by $W_{\mathrm{tree}}=\mathrm{Tr}\Phi^2$. This gauge theory is realized in type IIB string theory on $\mathcal{O}(-1)+\mathcal{O}(-1)$ bundle over $\mathbf{P^1}$ as the world volume theory on $N$ $D5$ branes wrap on the bare $\mathbf{P^1}$. We introduce another $D5$ branes wrap on holomorphic $2$-cycles not intersecting the $\mathbf{P^1}$ to include massive flavors. The massive flavors are engendered in strings stretching between these $D5$ branes and the $N$ $D5$ branes wrap on $\mathbf{P^1}$. 
  
Next we discuss the generalization to $U(N)$ gauge theories with a tree level superpotential,
\begin{eqnarray}
W_{\mathrm{tree}}=\sum_{p=1}^{n+1}\frac{g_p}{p}\mathrm{Tr}\Phi^p.
\end{eqnarray}
 $U(N)$ gauge theory with this tree level superpotential, but without massive flavors, was discussed in \cite{CIV}. In that paper, a Calabi-Yau manifold was generalized as:
\begin{eqnarray}
\left(\begin{array}{cc}
                      y+iz & w+iW_{\mathrm{tree}}^{\prime}(x) \\
                      -w+iW_{\mathrm{tree}}^{\prime}(x) & y-iz
               \end{array}
               \right) \left(\begin{array}{cc}
                      \lambda_1  \\
                      \lambda_2
               \end{array}
               \right) =0 ,
\end{eqnarray}
in $\mathbf{C}^4\times \mathbf{P^1}$. $\lambda_1$ and $\lambda_2$ are projective coordinates of $\mathbf{P^1}$ and $x,y,z,w$ are coordinates of $\mathbf{C}^4$. Note that there are ${\bf S^2}$'s at the points which satisfy $W^{\prime}_{\mathrm{tree}}(x)\equiv g_{n+1}\prod _{i=1}^n(x-a_i)=0$. $N$ $D5$ branes are distributed to these ${\bf S^2}$. If $N_i$ branes wrap on $i$-th ${\bf S^2}_i$, the gauge symmetry is broken as
\begin{eqnarray}
U(N)\to \prod_{i=1}^n U(N_i) \qquad \mathrm{with} \quad \sum_{i=1}^nN_i=N. \label{sas2} 
\end{eqnarray}
Again, to include massive flavors we introduce another $D5$ branes wrap on holomorphic $2$-cycles at points. The distance of that points from $N$ $D5$ branes are proportional to the mass scale of the massive flavors. Thereby, we obtain $4$ dimensional $\mathcal{N}=1$ $U(N)$ gauge theory with $N_f$ massive flavors.

Next we generalize the discussion above for $U(N)$ gauge theory to $SO(N)/Sp(N)$ gauge theories. As discussed in \cite{EOT}, we have to consider orientifold projection to realize $SO(N)/Sp(N)$ gauge theories. Under the orientifold projection, the coordinates introduced above are transformed as
\begin{equation}
(x,y,z,w,\lambda_1,\lambda_2) \to (-x,-y,-z,-w,\lambda_1,\lambda_2).
\label{complex conjugation} 
\end{equation}
The coordinates $\lambda_1$ and $\lambda_2$ are invariant under the projection; therefore, the orientifold
plane can wrap on $\mathbf{P^1}$ at $x=y=z=w=0$. 
We realize that the dimension of the orientifold plane is six, namely $O5$-plane, because the orientifold plane fills four dimensional space-time.
 In this geometry, world-volume theories on D5-branes are $SO/Sp$ gauge theories with the following tree level superpotential \cite{EOT}:
\begin{equation}
W_{SO/Sp}(\Phi)= \sum_{p=1}^{n+1}\frac{g_{2p}}{2p}{\rm Tr}\Phi^{2p}
\equiv\sum_{p=1}^{n+1}g_{2p}u_{2p}, \label{classical pote}
\end{equation}
where $\Phi$ is a chiral superfield in the adjoint representation 
of $SO(N)/Sp(N)$ gauge group and $u_{2p}\equiv\frac{1}{2p}{\rm Tr}\Phi^{2p}$. In distinction from $U(N)$ case, we use the notation $W_{SO/Sp}$ to represent the tree level superpotential. We define parameters $a_i$ by
\begin{eqnarray}
W^{\prime}_{SO/Sp}(x)=\sum_{p=1}^{n+1}g_{2p}x^{2p-1}
=g_{2n+2}\ x\prod_{i=1}^{n}(x^2+a_i^2). \label{classical sup} 
\end{eqnarray}
In a classical vacuum of these gauge theories, 
the eigenvalues of $\Phi$ become $0,\pm ia_i$, roots of $W^{\prime}_{SO/Sp}(x)=0$.
When $N_0$ D5-branes wrap on ${\bf S^2}$ at $x=0$ and 
$N_i$ D5-branes wrap on the ${\bf S^2}$ located at $x=\pm ia_i$, 
characteristic function of the gauge theories become, classically, 
$P(x)\equiv{\rm det}(x-\Phi)=x^{N_0}\prod_{i=1}^{n}(x^2+a_i^2)^{N_i}$
and the gauge groups break as,
\begin{eqnarray}
SO(N)\to SO(N_0)\times \prod_{i=1}^{n}U(N_i),\quad
Sp(N)\to Sp(N_0)\times \prod_{i=1}^{n}U(N_i), \label{sas1}
\end{eqnarray}
where $N=N_0+\sum_{i=1}^{n}2N_i$.

\subsection{Geometric dual description \label{generaleff}}
\subsubsection{No flavor case}
The dual descriptions of the gauge theories are found via geometric transition \cite{Vafa,CIV,EOT}. Under the transition each of the $\mathbf{S^2}_i$ on which $N_i$ $D5$ branes are wrapped have shrunk and have been replaced by the $\mathbf{S^3}_i$. After the transition the geometry corresponding to $U(N)$ gauge theory is given by 
\begin{eqnarray}
g\equiv W^{\prime}_{\mathrm{tree}}(x)^2+f_{n-1}(x)+y^2+z^2+v^2=0, \label{deformedCY} 
\end{eqnarray}
where $f_{n-1}(x)\equiv \sum_{i=0}^{n-1} b_i x^i$ is a degree $n-1$ th polynomial. In this deformed geometry, we choose a basis $A_i,B_i$ of $3$-cycles as symplectic pairing, $(A_i,B_j)=\delta_{ij}$.
These $3$-cycles are constructed as $\mathbf{S^2}$ fibration over the 
line segments between two critical points of ${W_{\mathrm{tree}}}^{\prime}(x)^2+f_{n-1}(x)$, 
$a_i^-, a_i^+$ and $\infty$ in $x$-plane.
Therefore we set the three cycles $A_i$ to be ${\bf S^2}$ fibration over the line segment between $a^-_i$ and $a^{+}_i$ and three cycles $B_i$ to be ${\bf S^2}$ fibration 
over the line segment between $a^+_i$ and $\Lambda_0$. Here we introduce cut-off $\Lambda_0$ because $B_i$ cycles are non-compact ones. 

The periods $S_i$ and dual periods $\Pi_i$ for 
this deformed geometry are given by the integrals of the holomorphic $3$-form $\Omega \equiv \frac{dxdydzdw}{dg}$:
\begin{equation}
S_i=\int_{A_i}\Omega, \quad \Pi_i=\int_{B_i}\Omega=\partial {\cal F}/\partial S_i.
\end{equation}
The dual periods are expressed in terms of the prepotential 
${\mathcal F}(S_i)$ of the deformed geometry. 
Because these $3$-cycles are constructed as ${\bf S^2}$ fibration, 
these periods are written in terms of the integrals over the $x$-plane as
\begin{eqnarray}
S_i=\frac{1}{2\pi i}\int_{a_i^{-}}^{a_i^{+}}\omega, \quad
\Pi_i=\frac{1}{2\pi i}\int_{a_i^{+}}^{\Lambda_0}\omega, \quad \omega=dx\ \left(W^{\prime}(x)^2+f_{n-1}(x)\right)^{\frac{1}{2}}, \label{fo10}
\end{eqnarray}
where $\omega$ is one form defined by the integral of $\Omega$ over the fiber ${\bf S^2}$. According to \cite{Vafa,CIV}, these periods $S_i$ are identified with the glueball superfields of $SU(N_i)$ gauge theories,
$S_i=\frac{1}{32\pi^2}{\rm Tr}_{SU(N_i)}W_{\alpha}W^{\alpha}$.

When geometric transition occurs, the $\mathbf{S^2}_i$ on which $N_i$ D$5$-branes wrap in the resolved geometry are replaced by RR $3$-form fluxes through the special Lagrangian $3$-cycles which we denote as $\mathbf{S^3}_i$ and NSNS $3$-form fluxes through dual cycles in the deformed geometry. These $3$-form fluxes generate a superpotential. In addition, ${\mathcal N}=2$ supersymmetry for the dual theory 
is broken partially to ${\mathcal N}=1$ supersymmetry \cite{Gukov,TV},
\begin{equation}
-\frac{1}{2\pi i}W_{{\rm eff}}
=\int\Omega\wedge(H_{{\rm R}}+\tau H_{{\rm NS}}),
\label{flux pot}
\end{equation}
where $H_{R}$ and $H_{NS}$ are $3$-form fluxes and $\tau$ is the
complexified Type IIB string coupling constant. In the dual theory defined by geometric transition, 
$H_{\rm R}$ and $H_{NS}$ have to satisfy the following relations:
\begin{eqnarray}
N_i=\int_{A_i}H_{{\rm R}}, \quad 
\alpha=\int_{B_i}\tau H_{{\rm NS}} \label{4maycon1},
\end{eqnarray}
where $\alpha$ is related to a bare gauge coupling constant $g_0$ of the $4$ dimensional $U(N)$ gauge theory through the relation $\alpha\equiv 4\pi i/g_0^2$. 

Substituting (\ref{4maycon1}) into (\ref{flux pot}), we find that the superpotential for the dual theory can be expressed in the following manner
in terms of periods $S_i$ and dual periods $\Pi_i$ of 
the deformed geometry,
\begin{equation}
-\frac{1}{2\pi i}W_{{\rm eff}}
=\sum_{i=1}^{n}N_i\Pi_i+\alpha\sum_{i=1}^{n}S_i. 
\label{superfulx}
\end{equation}

For $SO/Sp$ gauge theories the geometry after geometric transition is given by \cite{EOT}\begin{eqnarray}
W_{SO/Sp}^{\prime}(x)^2+f_{2n}(x)+y^2+z^2+w^2=0,
\end{eqnarray}
where $f_{2n}(x)$ is an even polynomial of degree $2n$. 
As in the $U(N)$ case, we can reduce this deformed geometry to a Riemann surface that has $2n+1$ branch cuts. We denote a holomorphic one-form on the Riemann surface for these cases as 
\begin{eqnarray}
\omega_{SO/Sp}=dx \left( W_{SO/Sp}^{\prime}(x)^2 +f_{2n}(x)  \right)^{\frac{1}{2}}.
\end{eqnarray}
Using this one-form we can represent the period integrals on the Riemann surface as follows:
\begin{eqnarray}
S_i=\frac{1}{2\pi i}\int_{ia_i^{-}}^{ia_i^{+}}\omega_{SO/Sp}, \quad
\Pi_i=\frac{1}{2\pi i}\int_{ia_i^{+}}^{\Lambda_0}\omega_{SO/Sp}, \ \ \ i=0,\cdots ,n .
\end{eqnarray}
The effective superpotential is given in the same way as the $U(N)$ case. After we reduce the general formula (\ref{flux pot}) written in $(x,y,z,w)$ to a formula on a Riemann surface written only in $(x,y)$, we can describe the effective superpotential in terms of $\omega_{SO/Sp}$ as
\begin{equation}
-\frac{1}{2\pi i}W_{{\rm eff}}
=\left(\frac{N_0}{2}\mp 1 \right)\Pi_0 + \sum_{i=1}^{n}N_i\Pi_i+\alpha\sum_{i=0}^{n}S_i , 
\label{SOSPsuperfulx}
\end{equation}
where the number of flux through $0$-th cut is reduced by the orientifold projection (\ref{complex conjugation}).

\subsubsection{Adding flavor}
Next we discuss gauge theories with $N_f$ massive fundamental matter multiplets. 
The effective superpotential for the $U(N)$ gauge theory with $W_{\mathrm{tree}}=\mathrm{Tr} \Phi^2$ was given in \cite{CIV}. We generalize results given in \cite{CIV} to gauge theories with an arbitrary polynomial tree level superpotential and classical gauge groups. The effective superpotential that comes from the contribution of flavors is given by the integral of $\omega$, and
\begin{eqnarray}
W_{\mathrm{eff}}^{flavor}=\frac{1}{2}\sum_{a=1}^{N_f}\int_{-m_a}^{\Lambda_0} \omega \equiv 2\pi i \sum_{a=1}^{N_f}F_a \label{addflux} ,
\end{eqnarray}
where $m_a$ is mass of the $a$-th flavor and where we choose the position of holomorphic $2$-cycles at $x=-m_a$. Because $x$ plane is the eigenvalue plane of $\Phi$, $x$ has the same dimension as mass. The reason for addition of these terms comes from the RR charge of the $D5$ brane wraps on holomorphic $2$-cycles that do not intersect with $\mathbf{S^2}_i$. These $2$-cycles do not shrink under geometric transition, so the geometry after the transition has the same number of cuts in the $x$-plane as a no-flavor case. However, RR-flux on the $x$-plane differs from the no-flavor case. $D5$ branes wrap on holomorphic $2$-cycles give a source for one unit of RR flux at $x=-m_a$. Therefore, integrals of $H_{R}$ around $x=-m_a$ indicate that
\begin{eqnarray}
\oint_{-m_a}H_{R}=1. \label{4maycon2}
\end{eqnarray}
We obtain the additional effective superpotential (\ref{addflux}) because the effective superpotential (\ref{flux pot}) is generated by fluxes.

The full effective superpotential is given as
\begin{eqnarray}
-\frac{1}{2\pi i}W_{\mathrm{eff}}=\sum_{i=1}^{n}N_i\Pi_i+\alpha\sum_{i=1}^{n}S_i-\sum_{a=1}^{N_f}F_a,
\end{eqnarray}
which depends on the cut-off parameter $\Lambda_0$, which comes from $\Pi_0,\Pi_i$ and the integral in (\ref{addflux}). 
From the monodromy argument, we can see the holomorphic beta function from a geometric viewpoint as follows. Under $\Lambda_0\to e^{2\pi i}\Lambda_0$, $\Pi_i$ and $F_a$ change as
\begin{eqnarray}
\Delta \Pi_i=-2 (\sum_{j=1}^{n}S_j ), \qquad \Delta F_a=-\sum_{j=1}^{n}S_j.
\end{eqnarray}
Factor two comes from the two copies of $x$-plane connected by branch cuts. However, $F_a$ does not have this factor because this integral comes from the one "semi 3-cycle''. Therefore, we see that $W_{\mathrm{eff}}$ must depend on the cutoff $\Lambda_0$ as 
\begin{eqnarray}
W_{\mathrm{eff}}&=&\cdots + 2\sum_{i=1}^{n}N_i \sum_{j=1}^{n}S_j\log \Lambda_0- \sum_{a=1}^{N_f}\sum_{j=1}^{n}S_j\log \Lambda_0+\alpha \sum_{j=1}^{n}S_j \nn \\
{}&=&\cdots +\left((2N-N_f) \log \Lambda_0+\alpha \right) \sum_{j=1}^{n}S_j , \label{5maycon1}
\end{eqnarray}
where $\cdots$ are single valued terms of $\Lambda_0$.
This $\log\Lambda_0$ piece can be absorbed into bare coupling constant $\alpha$. Let us introduce the new parameter $\Lambda$ and assume that
\begin{eqnarray}
\alpha=b_0 \log \frac{\Lambda}{\Lambda_0}.
\end{eqnarray}
We identify $\Lambda$ with the dynamically generated scale of $U(N)$ gauge theory and $b_0$ with a coefficient of one-loop holomorphic beta function. Inserting this relation into (\ref{5maycon1}), we see that $b_0$ must be the following value to cancel log divergent terms
\begin{eqnarray}
b_0=2N-N_f.
\end{eqnarray}
Note that this is a result obtained in terms of the geometric viewpoint and agrees with the beta function for $\mathcal{N}=1$ $U(N)$ gauge theory with $N_f$ flavors and one adjoint chiral matter \cite{peskin}. Therefore, this agreement provides partial justification for the dual geometry discussed in the previous section. We must study effective superpotential beyond monodoromic argument to get stronger justification. Before doing this we want to extend the discussion above to $SO/Sp$ gauge theories.

We can extend the discussion above for $U(N)$ gauge theory to the $SO/Sp$ gauge theories. For the $SO/Sp$ case, the deformed geometry has $Z_2$ identification for $x$ because both $W_{SO/Sp}(x)$ and $f_{2n}(x)$ are even functions of $x$. Therefore, considering the covering space, fluxes on the plane have a source at $x=\pm m_a$ in terms of contribution of $N_f$ flavors. The effective superpotential generated by these fluxes is given as
\begin{equation}
-\frac{1}{2\pi i}W_{{\rm eff}}
=\left(\frac{N_0}{2}\mp 1 \right)\Pi_0 + \sum_{i=1}^{n}N_i\Pi_i+\alpha\sum_{i=0}^{n}S_i-\sum_{a=1}^{N_f}F_a,
\label{flSOSPsuperfulx}
\end{equation}
where the period integrals $S_i$ and $\Pi_i$ are defined in terms of $\omega_{SO/Sp}$. Similarly to the $U(N)$ case, we can obtain coefficients of holomorphic beta functions for $SO/Sp$ gauge theories from monodromy argument as:
\begin{eqnarray}
b_0&=&2(N-2)-2N_f \qquad    \mathrm{for}\ \ SO(N) \\
b_0&=&2(N+2)-2N_f  \qquad  \mathrm{for}\ \ Sp(N).
\end{eqnarray}
These results agree with the beta functions for ${\cal N}=1$ $SO/Sp$ gauge theories with $N_f$ flavors and adjoint chiral matter \cite{peskin}. The proceeding discussion uses identical notation $b_0$ to the coefficients of holomorphic beta functions for $U(N)$ and $SO(N)/Sp(N)$ gauge theories.

\section{Effective Superpotential}
This section provides a proof of the equivalence of two potentials $W_{\mathrm{eff}}$ and $W_{\mathrm{low}}$. The effective superpotential $W_{\mathrm{eff}}$ is derived from the Calabi-Yau geometry with fluxes. On the other hand, $W_{\mathrm{low}}$ is the superpotential derived purely from  field theory analysis. First, we discuss $U(N)$ gauge theory and then $SO(2N)$ gauge theory. For $SO(2N+1)$ and $Sp(2N)$ theories, we do not explicitly address $SO(2N)$ gauge theory. We merely provide some comments because the discussions are very similar.

\subsection{Field theory analysis}
We concentrate on the Coulomb branch, in which the adjoint chiral superfield $\Phi$ has following expectation values in the classical limit $\Lambda \to 0$, 
\begin{eqnarray}
P(x)&\equiv& \langle \mathrm{det}(x-\Phi) \rangle \to \prod_{i=1}^n (x-a_i)^{N_i} \label{clvac} \qquad \mathrm{for} \ \ U(N) \\
P(x)&\equiv& \langle \mathrm{det}(x-\Phi) \rangle \to x^{N_0}\prod_{i=1}^n (x^2+a_i^2)^{N_i} \label{sospclvac} \qquad \mathrm{for} \ \ SO/Sp ,
\end{eqnarray}
where $a_i$ are roots of $W^{\prime}_{\mathrm{tree}}(x)=0$ or $W_{SO/Sp}^{\prime}(x)=0$. These classical vacua have unbroken gauge groups as (\ref{sas2}) or (\ref{sas1}).
Next we briefly review the Seiberg-Witten curves for $U(N)$ and $SO(N)/Sp(N)$ gauge theories with $N_f$ flavors given in \cite{Tera9101112} because we will discuss field theory analysis for effective superpotential using the Seiberg-Witten curve with monopole massless constraint:
\begin{eqnarray}
y^2&=&P_N(x)^2-4 \Lambda^{2N-N_f}\prod_{a=1}^{N_f}(x+m_a),\qquad \qquad \mathrm{for} \ \ U(N) \label{SWUN3may} \\
y^2&=&P_{2N}(x)^2-4 \Lambda^{4N-2-2N_f}x^2\prod_{a=1}^{N_f}(x^2-m_a^2), \qquad \mathrm{for} \ \ SO(2N+1) \\
y^2&=&P_{2N}(x)^2-4 \Lambda^{4N-4-2N_f}x^4\prod_{a=1}^{N_f}(x^2-m_a^2), \qquad \mathrm{for} \ \ SO(2N) \label{SWcurveSO}.
\end{eqnarray}
The curve for $Sp(2N)$ theory differs slightly from the ones for the other gauge groups,
\begin{eqnarray}
y^2=\left(x^2 P_{2N}(x)+2\Lambda^{2N+2-N_f}\prod_{a=1}^{N_f}m_a \right)^2-4 \Lambda^{2(2N+2-N_f)}\prod_{a=1}^{N_f}(x^2-m_a^2). \label{SWcurveSp}
\end{eqnarray}
Characteristic functions $P(x)$ are different for each classical group because they are defined in terms of adjoint matter multiplets $\Phi$ as in (\ref{clvac}). Even though  $P(x)$ is an $N$th-order polynomial for the $U(N)$ case, they become $2N$th order polynomials for $SO(2N)$, $SO(2N+1)$, and $Sp(2N)$ cases.

As in \cite{CIV}, a supersymmetric vacuum with $U(1)^n$ unbroken gauge groups necessarily has at least $l=N-n$ mutually local massless monopoles. Therefore, we consider singular points or locus in the moduli space where $l=N-n$ mutually local monopoles become massless. This means that $l$ one cycles shrink to zero; therefore, the Seiberg-Witten curve has $l$ double roots:
\begin{eqnarray}
y^2=P_N(x)^2-4\Lambda^{2N-N_f}\prod_{a=1}^{N_f}(x+m_a)=H_{l}^2(x)F_{2N-2l}(x)\equiv \prod_{i=1}^{l}(x-p_i)^2F_{2N-2l}(x). \label{fieldmono}
\end{eqnarray}
Regarding the $SO/Sp$ case, because the Seiberg-Witten curves are even functions, the massless monopole constraint becomes the following:
\begin{eqnarray}
y^2&=&x^2H_{2(l-1)}^2(x)F_{4N-4l+2}(x)\equiv  x^2 \prod_{i=1}^{l-1} (x^2-p_i^2)^2F_{4N-4l+2}(x) \ \ \ \mathrm{for} \ \ SO(N) \label{masslessSO} \\
y^2&=&x^2H_{2l}^2(x)F_{4N-4l+2}(x)\equiv  x^2 \prod_{i=1}^l (x^2-p_i^2)^2F_{4N-4l+2}(x) \ \ \ \mathrm{for} \ \ Sp(2N) .
\end{eqnarray}
There are $l$ double roots in the region $x\ge 0$. Because $y^2$ for $SO/Sp$ theories are even functions of $x$, $F_{4N-4l+2}(x)$ is also even function.

\subsubsection{U(N) case}
Here we concentrate on the $U(N)$ gauge theory. The low energy superpotential with the constraint (\ref{fieldmono}) is described as 
\begin{eqnarray}
W_{\mathrm{low}}=\sum_{r=1}^{n+1}g_r u_{r}+\sum_{i=1}^{l} \left[L_i \left(P_N(p_i)-2\epsilon_i \Lambda^{\frac{b_0}{2}}\sqrt{A(p_i)}\right) +Q_i \frac{\partial }{\partial p_i}\left( P(p_i)-2\epsilon_i \Lambda^{\frac{b_0}{2}}\sqrt{A(p_i)}  \right) \right] .
\end{eqnarray}
Therein, $L_i,Q_i$ are Lagrange multipliers imposing the condition (\ref{fieldmono}). In addition, we define two new notations: $\epsilon_i=\pm 1$, $A(x)\equiv \prod_{a=1}^{N_f}(x+m_a)$. We obtain the following equations from the equations of motion for $p_i$ and $Q_i$:
\begin{eqnarray}
Q_i=0, \qquad \frac{\partial }{\partial p_i}\left( P_N(p_i)-2\epsilon_i \Lambda^{\frac{b_0}{2}}\sqrt{A(p_i)}\right)=0.
\end{eqnarray}
Therefore we can use $Q_i=0$ at the level of equation of motion. 
The variation of $W_{\mathrm{low}}$ with respect to $u_r$ leads to 
\begin{eqnarray}
g_r+\sum_{i=1}^{l}L_i\frac{\partial}{\partial u_r}\left(P_N(p_i)-2\epsilon_i \Lambda^{\frac{b_0}{2}}\sqrt{A(p_i)}\right)=0.
\end{eqnarray}
The third term vanishes because $A(p_i)$ is independent of $u_r$. Using $P_N(x)=\sum_{k=0}^Nx^{N-k}s_k$, we can express the coefficients of tree level superpotential as 
\begin{eqnarray}
g_r=\sum_{i=1}^{l}\sum_{j=0}^{N}L_ip_i^{N-j}s_{j-r}. 
\end{eqnarray}
With this relation, as in \cite{DO,Ahn,TY}, we can obtain the following relation:
\begin{eqnarray}
W^{\prime}_{\mathrm{tree}}=\sum_{r=1}^{N}g_rx^{r-1}&=&\sum_{r=-\infty}^{N} \sum_{i=1}^{l} \sum_{j=0}^{N} x^{r-1}p_i^{N-j}s_{j-r}L_i-x^{-1}\sum_{i=1}^{l} L_i P_N(p_i)+\mathcal{O}(x^{-2}) \nn \\
{}&=&\sum_{r=-\infty}^{N} \sum_{i=1}^{l} \sum_{j=0}^{N} x^{r-1}p_i^{N-j}s_{j-r}L_i-x^{-1}\sum_{i=1}^{l}2\epsilon_i L_i  \Lambda^{\frac{b_0}{2}}\sqrt{A(p_i)} +\mathcal{O}(x^{-2}) \nn \\
{}&=&\sum_{j=-\infty}^{N} \sum_{i=1}^{l} P_N(x) x^{j-N-1}L_i p_i^{N-j}-x^{-1}\sum_{i=1}^{l}2\epsilon_i L_i  \Lambda^{\frac{b_0}{2}}\sqrt{A(p_i)} +\mathcal{O}(x^{-2}) \nn \\
{}&=&\sum_{i=1}^{l} \frac{P_N(x)}{x-p_i} L_i-x^{-1}\sum_{i=1}^{l}2 \epsilon_i L_i \Lambda^{\frac{b_0}{2}}\sqrt{A(p_i)} +\mathcal{O}(x^{-2}) \label{3maycon1} .
\end{eqnarray}
As in \cite{CIV}, we define new $l-1$-th polynomial $B_{l-1}$ as follows: 
\begin{eqnarray}
\sum_{i=1}^{l} \frac{L_i}{x-p_i}=\frac{B_{l-1}(x)}{H_l(x)}.
\end{eqnarray}
After inserting this equation into (\ref{3maycon1}) and using the Seiberg-Witten curve (\ref{SWUN3may}), we obtain the following relation: 
\begin{eqnarray}
W_{\mathrm{tree}}^{\prime}(x)+x^{-1}\sum_{i=1}^l2\epsilon_i L_i \Lambda_0^{\frac{b_0}{2}}\sqrt{A(p_i)}=B_{l-1}(x)\sqrt{F_{2N-2l}(x)-\frac{4\Lambda^{b_0}A(x)}{H_l(x)}}+\mathcal{O}(x^{-2}) \label{3maycon2} .
\end{eqnarray}
We infer that $B_{l-1}(x)$ should be on the order of $n-N+l$ because the highest order terms in $W_{\mathrm{tree}}^{\prime}(x)$ are $g_{n+1}x^n$. This indicates that $l\ge N-n$; in particular, $B_{N-n-1}=g_{n+1}$ is constant for $l=N-n$. Using this relation, we can represent (\ref{3maycon2}) as
\begin{eqnarray}
g_{n+1}^2F_{2n}(x)={W_{\mathrm{tree}}^{\prime}}^2+4g_{n+1}x^{n-1}\Lambda^{\frac{b_0}{2}}\sum_{i=1}^{l}\epsilon_i L_i \sqrt{A(p_i)}+\mathcal{O}(x^{n-2}) ={W^{\prime}_{\mathrm{tree}}}^2+f_{n-1}(x). \label{3maycon3} 
\end{eqnarray}
Substituting this relation, we can represent the constraint (\ref{fieldmono}) as \begin{eqnarray}
P_N(x)^2-4\Lambda^{b_0}A(x)=\frac{1}{g_{n+1}^2}\left(W^{\prime}_{\mathrm{tree}}(x)^2+f_{n-1}(x) \right)H_{N-n}^2(x). \label{fieldnewmono}
\end{eqnarray}
Therefore, remembering the definition $f_{n-1}(x)\equiv \sum_{k=0}^{n-1} b_k x^k$, from (\ref{3maycon3}) we can read off deformation parameter $b_{n-1}$ as \begin{eqnarray}
b_{n-1}=4g_{n+1}x^{n-1}\Lambda^{\frac{b_0}{2}}\sum_{i=1}^{l}\epsilon_i L_i \sqrt{A(p_i)} \label{bn1}.
\end{eqnarray}
On the other hand, the derivative of $W_{\mathrm{low}}$ with respect to $\Lambda$ is written as
\begin{eqnarray}
\frac{\partial W_{\mathrm{low}}}{\partial \log \Lambda^{b_0}}=-\frac{1}{4g_{n+1}} 4g_{n+1}x^{n-1}\Lambda^{\frac{b_0}{2}}\sum_{i=1}^l \epsilon_i L_i \sqrt{A(p_i)}=-\frac{1}{4g_{n+1}} b_{n-1}, \label{fieldeffb}
\end{eqnarray}
where we use (\ref{bn1}) in the last equality. This is a salient result for proof of the equivalence $W_{\mathrm{low}}$ and $W_{\mathrm{eff}}$.

Next we consider the superpotential $W_{\mathrm{low}}$ in the classical limit $\Lambda \to 0$. Because we are considering the classical vacuum (\ref{clvac}), the classical value of effective superpotential is given as
\begin{eqnarray}
W_{\mathrm{low}}(\Lambda \to 0)=\sum_{i=1}^n N_i \sum_{k=1}^{n+1}\frac{1}{k}g_ka_i^k. \label{fieldeffcl}
\end{eqnarray}
We will show a justification for geometric dual description introduced in the previous section using results of (\ref{fieldeffb}) and (\ref{fieldeffcl}). However, prior to this, we extend the above discussion to $SO/Sp$ gauge theories.

\subsubsection{SO(2N) case}
The deformation function is an even function \cite{EOT,FO} for the $SO/Sp$ case. With this in mind, we can proceed the discussion in the similar way as $U(N)$ case. We address only $SO(2N)$ gauge theories in detail and only give results for the others. From the equation (\ref{masslessSO}), we have double roots at $x=0,\pm p_i \ i=1,\cdots, l-1$. However, we do not need to consider all points. We only need to consider $x=0$ and $x=+p_i$ because $P_{2N}(x)$ and $\widetilde{A}(x)\equiv x^4 \prod_{a=1}^{N_f}(x^2-m_a^2)$ are even functions of $x$. For simplicity of equations, we define $p_l=0$ such that index $i$ runs from $1$ to $l$. By these conventions, we have effective superpotential with massless monopole constraints (\ref{masslessSO}),
\begin{eqnarray}
W_{\mathrm{low}}=\sum_{r=1}^{n+1}g_{2r} u_{2r}+\sum_{i=1}^{l} \left[L_i \left(P_{2N}(p_i)-2\epsilon_i \Lambda^{\frac{b_0}{2}}\sqrt{\widetilde{A}(p_i)}\right) +Q_i \frac{\partial }{\partial p_i}\left( P_{2N}(p_i)-2\epsilon_i \Lambda^{\frac{b_0}{2}}\sqrt{\widetilde{A}(p_i)}  \right) \right] \nn ,
\end{eqnarray}
where $L_i,Q_i$ are Lagrange multipliers imposing the condition (\ref{masslessSO}) and $\epsilon_i=\pm 1$. From the equation of motion for $p_i$ and $Q_i$, we obtain
\begin{eqnarray}
Q_i=0, \qquad \frac{\partial }{\partial p_i}\left( P_{2N}(p_i)-2\epsilon_i \Lambda^{\frac{b_0}{2}}\sqrt{\widetilde{A}(p_i)}\right)=0.
\end{eqnarray}
The variation of $W_{\mathrm{low}}$ with respect to $u_{2r}$ engenders
\begin{eqnarray}
g_{2r}+\sum_{i=1}^{l}L_i\frac{\partial}{\partial u_{2r}}\left(P_{2N}(p_i)-2\epsilon_i \Lambda^{\frac{b_0}{2}}\sqrt{\widetilde{A}(p_i)}\right)=0.
\end{eqnarray}
Because $\widetilde{A}(p_i)$ is independent of $u_{2r}$, the third term vanishes. Using $P_{2N}(x)=\sum_{j=0}^{N}s_{2j}x^{2N-2j}$, we obtain 
\begin{eqnarray}
g_{2r}=\sum_{i=1}^{l}\sum_{j=0}^{N}L_ip_i^{2N-2j}s_{2j-2r}. 
\end{eqnarray}
With this relation, as in \cite{DO,Ahn,TY}, we obtain the following relation:
\begin{eqnarray}
W^{\prime}_{SO/Sp}&=&\sum_{r=1}^{N}g_{2r}x^{2r-1} \nn \\
&=&\sum_{r=-\infty}^{N} \sum_{i=1}^{l} \sum_{j=0}^{N} x^{2r-1}p_i^{2N-2j}s_{2j-2r}L_i-x^{-1}\sum_{i=1}^{l} L_i P_{2N}(p_i)+\mathcal{O}(x^{-3}) \nn \\
{}&=&\sum_{r=-\infty}^{N} \sum_{i=1}^{l} \sum_{j=0}^{N} x^{2r-1}p_i^{2N-2j}s_{2j-2r}L_i-x^{-1}\sum_{i=1}^{l}2\epsilon_i L_i  \Lambda^{\frac{b_0}{2}}\sqrt{\widetilde{A}(p_i)} +\mathcal{O}(x^{-3}) \nn \\
{}&=&\sum_{j=-\infty}^{N} \sum_{i=1}^{l} P_{2N}(x) x^{2j-2N-1}L_i p_{i}^{2N-2j}-x^{-1}\sum_{i=1}^{l}2\epsilon_i L_i  \Lambda^{\frac{b_0}{2}}\sqrt{\widetilde{A}(p_i)} +\mathcal{O}(x^{-3}) \nn \\
{}&=&\sum_{i=1}^{l} \frac{xP_{2N}(x)}{x^2-p_i^2} L_i-x^{-1}\sum_{i=1}^{l}2\epsilon_i L_i \Lambda^{\frac{b_0}{2}}\sqrt{\widetilde{A}(p_i)} +\mathcal{O}(x^{-3}) \label{3maycond3} .
\end{eqnarray}
Defining $B_{2(l-1)}$ as in \cite{CIV}, 
\begin{eqnarray}
\sum_{i=1}^{l} \frac{L_i}{x^2-p_i^2}=\frac{B_{2(l-1)}(x)}{x^2H_{2(l-1)}(x)}.
\end{eqnarray}
Thereby, after we insert this equation into (\ref{3maycon3}) and take into account the Seiberg-Witten curve (\ref{SWcurveSO}) and $B_{2N-2n-2}=g_{2n+2}$ for $l=N-n$, (\ref{3maycon3}) can be written as
\begin{eqnarray}
g_{2n+2}^2F_{4n+2}={W_{SO/Sp}^{\prime 2}}+4g_{2n+2}x^{2n}\Lambda^{\frac{b_0}{2}}\sum_{i=1}^{l}\epsilon_i L_i \sqrt{\widetilde{A}(p_i)}+\mathcal{O}(x^{2n-2}) ={W^{\prime 2}_{SO/Sp}} .+f_{2n}(x) \nn
\end{eqnarray}
Substituting this relation, we can represent the constraint (\ref{fieldmono}) as \begin{eqnarray}
P_{2N}(x)^2-4\Lambda^{b_0}\widetilde{A}(x)=\frac{1}{g_{2n+2}^2}\left(W^{\prime 2}_{SO/Sp}(x)+f_{2n}(x) \right)H_{2N-2n-2}^2(x). \label{sofieldnewmono}
\end{eqnarray}
Thus deformation parameter $b_{2n}$ is represented as
\begin{eqnarray}
b_{2n}=4g_{2n+2}\Lambda^{\frac{b_0}{2}}\sum_{i=1}^l\epsilon_i L_i \sqrt{\widetilde{A}(p_i)} .
\end{eqnarray}
Using this result, we obtain
\begin{eqnarray}
\frac{\partial W_{\mathrm{low}}}{\partial \log \Lambda^{b_0}}=-\frac{1}{4g_{2n+2}} b_{2n}. \label{5maycond2}
\end{eqnarray}
Next we consider superpotential $W_{\mathrm{low}}$ in the classical limit $\Lambda \to 0$. Because we are assuming the classical vacua (\ref{sospclvac}), the classical value of the effective superpotential becomes
\begin{eqnarray}
W_{\mathrm{low}}(\Lambda \to 0)=2\sum_{i=1}^nN_i\sum_{k=1}^{n+1}\frac{g_{2k}}{2k}a_i^{2k}. \label{5maycond1}
\end{eqnarray}
In terms of these two results (\ref{5maycond2}) and (\ref{5maycond1}) in purely field theory analysis, we can prove the equivalence $W_{\mathrm{low}}$ and $W_{\mathrm{eff}}$ in the next subsection.

We can continue analysis for $SO(2N+1)$ and $Sp(2N)$ similarly to the $SO(2N)$ case. However $b_0$ becomes $4N-2-2N_f$ and $4N+2-2N_f$, respectively. Massless monopole constraints for these cases can be rewritten as
\begin{eqnarray}
P_{2N}(x)^2-4\Lambda^{b_0}x^2\prod_{a=1}^{N_f}(x^2-m^2_a)=\frac{1}{g_{2n+2}^2}\left(W^{\prime}_{SO/Sp}(x)^2+f_{2n} \right)H_{2N-2n-2}^2(x), \label{somono}
\end{eqnarray}
\begin{eqnarray}
\biggl(x^2 P_{2N}(x)+2\Lambda^{b_0}\prod_{a=1}^{N_f}m_a \biggr)^2\!\!-4 \Lambda^{2b_0}\prod_{a=1}^{N_f}(x^2-m_a^2)=\frac{1}{g_{2n+2}^2}\left(W^{\prime}_{SO/Sp}(x)^2+f_{2n} \right)H_{2N-2n}^2(x). \label{spmono}
\end{eqnarray}

\subsection{Geometric dual analysis \label{sub3.2}}
In this section we study the effective superpotential $W_{\mathrm{eff}}$ (\ref{flSOSPsuperfulx}) which comes from fluxes on the Calabi-Yau geometry and prove its equivalence to $W_{\mathrm{low}}$. As in the previous subsection, we consider the derivative of effective superpotential with respect to the $\Lambda$ and its value in the classical limit $\Lambda \to 0$. As in \cite{CV}, it is convenient to use deformation parameters $\{b_{n-1},\cdots, b_0 \}$ as a change of variables instead of periods $\{S_1, \cdots S_n \}$. The expectation values of $b_k$ are given by solving $\partial W_{\mathrm{eff}}/\partial b_k=0$, which minimize the effective superpotential. 

First we consider the derivative of $W_{\mathrm{eff}}$ with respect to $\Lambda$ for $U(N)$ gauge theory. The $\Lambda$ dependence of $W_{\mathrm{eff}}$ arises from two terms: $\Pi_i$ and $F_a$. Thus we obtain
\begin{eqnarray}
\frac{\partial W_{\mathrm{eff}}(\langle b_k\rangle ,\log \Lambda)}{\partial \log \Lambda^{2N-N_f}}&=&\sum_{i=1}^{n}N_i \frac{\partial \Pi_i}{\partial \log \Lambda^{2N-N_f}}-\sum_{a=1}^{N_f}\frac{\partial F_a}{\partial \log \Lambda^{2N-N_f}} \nn \\
{}&=&\sum_{i=1}^{n}S_i=-\frac{1}{4g_{n+1}}b_{n-1}. \label{effbb}
\end{eqnarray}
The Riemann surface that we are considering has two special points located at the two pre-images of infinity. We call these two points $P$ and $Q$. The sum of period $S_i$ becomes the integral of $\omega$ around $P$ or $Q$ because the integral of $\omega$ around $m_a$ is zero. This integral around $P$ or $Q$ gives $b_{n-1}$ for the residue. We have used the residue in the last equality of (\ref{effbb}).

Next we consider the effective superpotential in the classical limit $\Lambda\to 0$. We obtain following result under the limit because deformation function $f_{n-1}(x)$ becomes zero:
\begin{eqnarray}
W_{\mathrm{eff}}(\Lambda \to 0)=\sum_{i=1}^n N_i \sum_{k=1}^{n+1}\frac{1}{k}g_ka_i^k-\left(N-\frac{N_f}{2}\right)W_{\mathrm{tree}}(\Lambda_0)-\frac{1}{2}\sum_{a=1}^{N_f} W_{\mathrm{tree}}(-m_a). \label{classeff}
\end{eqnarray}
Since we could add an arbitrary constant term to $W_{\mathrm{eff}}$ as discussed in \cite{CV}, we can redefine the effective superpotential for dual geometry as
\begin{eqnarray}
W_{\mathrm{eff}}=-\sum_{i=1}^{n}N_i \int_{a^+_i}^{\Lambda_0} \omega-\alpha \sum_{i=1}^{n}\int_{a^-_i}^{a^+_i}\omega+\frac{1}{2}\sum_{a=1}^{N_f} \int_{-m_a}^{\Lambda_0} \omega+\left(N-\frac{N_f}{2}\right)W_{\mathrm{tree}}(\Lambda_0)+\frac{1}{2}\sum_{a=1}^{N_f} W_{\mathrm{tree}}(-m_a). \nn
\end{eqnarray}
Here we added the last two terms. Then we obtain the following classical limit,
\begin{eqnarray}
W_{\mathrm{eff}}(\Lambda \to 0)=\sum_{i=1}^n N_i \sum_{k=1}^{n+1}\frac{1}{k}g_ka_i^k. \label{classeff2}
\end{eqnarray}
These two results (\ref{effbb}) and (\ref{classeff2}) agree with (\ref{fieldeffb}) and (\ref{fieldeffcl}) obtained from field theory. This agreement provides justification for the dual geometry introduced in section 2. 

We can obtain similar results for $SO/Sp$ gauge theories. We use variables $\{b_{2n}, \cdots ,b_{0}\}$ instead of periods $\{S_0,\cdots ,S_n\}$. The derivative of $W_{\mathrm{eff}}$ with respect to $\Lambda$ leads to
\begin{eqnarray}
\frac{\partial W_{\mathrm{eff}}(\langle b_{2k}\rangle ,\log \Lambda)}{\partial \log \Lambda^{b_0}}=\sum_{i=1}^{n}S_i=-\frac{1}{4g_{2n+2}}b_{2n}, \label{effb}
\end{eqnarray}
where we evaluate the period integral around $P$ and obtain the residue $b_{2n}$. After the addition of constant terms, the classical limit $\Lambda \to0$ engenders the result for the $SO/Sp$ case:
\begin{eqnarray}
W_{\mathrm{eff}}(\Lambda \to 0)=2\sum_{i=1}^nN_i\sum_{k=1}^{n+1}\frac{g_{2k}}{2k}a_i^{2k}.
\end{eqnarray}
Note that the number $N_0$ does not appear because $a_0=0$. These results agree with field theory results.

\section{Flux on Riemann Surface}
In this section we comment on the flux on the Riemann surface in the case with flavor, which has crucial difference from the case without flavors. In subsection \ref{generaleff}, we have discussed the reduction of period integrals on the certain cycles in Calabi-Yau manifold to integrals on a Riemann surface. Let us introduce one form $h$ obtained from the reduction of the three form $H$ to the Riemann surface:
\begin{eqnarray}
h=\int_{S^2}H, \qquad H=H_{R}+\tau H_{NS} \label{6maycon1} .
\end{eqnarray}
We can represent the conditions for fluxes on the Riemann surface using this one form $h$. We discuss the $U(N)$ case first, then generalize to the $SO/Sp$ case. From (\ref{4maycon1}) and (\ref{4maycon2}) the conditions are given as
\begin{eqnarray}
\int_{a_i^-}^{a_i^+}h=N_i,\qquad \int_{a_i^+}^{\Lambda_0}h =\tau_{YM},\qquad \oint_{-m_a}h=1. \label{fluxcond}
\end{eqnarray}
Let us introduce parameters $s$ and $t$ satisfying $0 \le s,t \le 1$, which are the relative numbers of flavor on the upper and lower sheets of the $x$-plane, respectively. Accordingly, we obtain further conditions for $h$,\begin{eqnarray}
\oint_{P}h=-N-sN_f,\qquad \oint_{Q}h=N-tN_f. \label{fluxPQ}
\end{eqnarray}
These conditions indicate that the one form $h$ on the Riemann surface should have simple poles at $-m_a$, $P$, and $Q$. Residues of these ploes are $1$, $-N-sN_f$, and $N-tN_f$, respectively. 

The equation of motion for $b_k$, $\partial W_{\mathrm{eff}}/\partial b_{k}=0$, suggests the following equation 
\begin{eqnarray}
N\int_Q^P \eta_k+\sum_{a=1}^{sN_f}\int_{P}^{-m_a} \eta_k +\sum_{b=1}^{tN_f} \int_{Q}^{-m_b} \eta_k=0,
\end{eqnarray}
where $\eta_k$ is a holomorphic one form defined by $\eta_k \equiv \frac{\partial \omega}{\partial b_k}$. Due to Abel's theorem, this equation implies that there is a meromorphic function with the divisor \cite{CV}:
\begin{eqnarray}
\sum_{a=1}^{N_f}(-m_a)-(N+sN_f)P+(N-tN_f)Q.
\end{eqnarray}
For simplicity, we assume $s=0$ and $t=1$, which means that all singularities representing a source of flux exist only on the lower sheet. We can describe the meromorphic function explicitly for $2N>N_f$. As discussed in \cite{CV} we introduce a new function $z$ defined by
\begin{eqnarray}
z= P_N(x)-\frac{1}{g_{n+1}}\sqrt{{W_{\mathrm{tree}}^{\prime 2}+f_{n-1}}}H_{N-n}(x) =P_N(x)-\sqrt{P_N(x)^2-4\Lambda^{2N-N_f} {\prod}_{a=1}^{N_f}(x+m_a)},
\end{eqnarray}
where we use (\ref{fieldnewmono}) at the last equality. This new function has an $N$th order pole at $P$ and an $N-N_f$th order zero at $Q$ and a $1$th order zero at $-m_a$. As in \cite{CV}, in terms of this function $z$ the flux $h$ can be written as
\begin{eqnarray}
h=-\frac{1}{2\pi i} \frac{dz}{z}. \label{oneform}
\end{eqnarray}
In fact this form has poles of order $1$ at $-m_a$, $Q$, and $P$ with residues $1$, $N-N_f$, and $-N$. It then gives expected relations (\ref{fluxcond}) and (\ref{fluxPQ}) for cases $s=0$ and $t=1$. 

For the $SO/Sp$ case, some of the conditions of fluxes, namely the integrals of $h$ on the cut, change as
\begin{eqnarray}
\int_{a_0^-}^{a_0^+}h =\frac{N_0}{2}\mp 1, \ \ \int_{ia_i^-}^{ia_i^+}h =N_i\ \ i=1, \cdots, n.
\end{eqnarray}
The equation of motion for $b_{2k}$ implies that
\begin{eqnarray}
\left(N\mp 2 \right)\int_Q^P \widetilde{\eta}_{2k}+2\sum_{a=1}^{sN_f}\int_{P}^{-m_a}\widetilde{\eta}_{2k}+2\sum_{b=1}^{tN_f}\int_Q^{-m_b}\widetilde{\eta}_{2k}=0,
\end{eqnarray}
where $\widetilde{\eta}_{2k}\equiv \frac{\partial \omega_{SO/Sp}}{\partial b_{2k}}$. For the $SO/Sp$ cases, we define function $z$ by
\begin{eqnarray}
z&=&P_{2N}(x)/x^2 -\sqrt{\left(P_{2N}(x)/x^2 \right)^2-4\Lambda^{4N+4-2N_f}{\prod}_{a=1}^{N_f}(x^2-m_a^2)},\quad \mathrm{for} \quad SO(2N) \\
z&=&B(x) -\sqrt{B(x)^2-4\Lambda^{2(2N+2-N_f)}{\prod}_{a=1}^{N_f}(x^2-m_a^2)},\quad \  \qquad \ \ \qquad \mathrm{for} \quad Sp(2N)
\end{eqnarray}
where $P_{2N}(x)$ is a $2N$th order polynomial and $B(x)\equiv x^2P_{2N}(x)+2\Lambda^{2N+2-N_f}\prod_{a=1}^{N_f}m_a$. These functions have $2N\mp 2 $th order pole at $P$ and $2N\mp 2-2N_f$th order zero at $Q$ and 1th order zero at $\pm m_a$ for $2(2N\mp 2)>2N_f$. In these cases, one-form flux is also given by (\ref{oneform}) and has a pole of order $1$ at $\pm m_a$, $Q$, and $P$ with residue $1$, $(2N\mp 2)-2N_f$, and $-(2N\mp 2)$.

\section{Affleck-Dine-Seiberg from Calabi-Yau with Flux}
In this section, we specifically address the gauge theories with quadratic superpotential $W_{\mathrm{tree}}=\frac{1}{2}\mathrm{Tr}\Phi^2$ or $W_{SO/Sp}=\frac{1}{2}\mathrm{Tr}\Phi^2$, where we choose $\frac{1}{2}$ for the coefficient of $\Phi^2$ for simplicity. In addition, we assume that all masses of flavors take the same value. Under these assumptions we can reproduce an Affleck-Dine-Seiberg superpotential \cite{ADS} from a dual geometrical viewpoint. In this simple case, the dual geometry (\ref{deformedCY}) and the period integral (\ref{fo10}) become
\begin{eqnarray}
f=x^2+y^2+z^2+v^2-\mu=0, 
\end{eqnarray}
\begin{eqnarray}
S=\frac{1}{2\pi i}\int_{-\sqrt{\mu}}^{\sqrt{\mu}}dx \sqrt{x^2-\mu}=\frac{\mu}{4},
\end{eqnarray}
respectively. Here we ignore the subscript for period integral $S$ because the double cover $x$-plane has only one branch cut. After computation of $\Pi$, we obtain the effective superpotential
\begin{eqnarray}
W_{\mathrm{eff}}=S \log \left(\frac{\Lambda^{2\hat{N}}}{S^{\hat{N}}} \right)+\hat{N}S+W_{\mathrm{eff}}^{flavor} |_{\Lambda_0\to \Lambda} ,
\end{eqnarray}
where we use $\hat{N}$ which stands for $N$ for $U(N)$ case and $N\mp 2$ for $SO(N)/Sp(N)$ cases, respectively. This $\pm 2$ term for $SO/Sp$ cases comes from a contribution of the orientifold plane \cite{SV,EOT,ACHKR}. The contribution from flavor leads to
\begin{eqnarray}
W_{\mathrm{eff}}^{flavor}&=&-\frac{1}{2}\sum_{a=1}^{N_f}\int_{-m}^{\Lambda_0}dx \sqrt{x^2-\mu} \nn \\
{}&=&N_f \left(-\frac{S}{2}-\frac{m^2}{4}\sqrt{1-\frac{4S}{m^2}}+S \log \frac{m}{\Lambda_0}+S\log \left(\frac{1}{2}+\frac{1}{2}\sqrt{1-\frac{4S}{m^2}} \right) \right). \label{Aruresult}
\end{eqnarray}
As discussed in subsection \ref{generaleff}, we can substitute $\Lambda$ for $\Lambda_0$; the former is interpreted as the dynamically generated energy scale because the $\log \Lambda_0$ divergent piece can be absorbed into bare coupling constant $\alpha$. Then, collecting $\log$ terms, we can express these terms as
\begin{eqnarray}
S\log \left(\frac{\Lambda^{2\hat{N}}}{S^{\hat{N}}} \right)+N_f S\log \frac{m}{\Lambda}=S \log \left(\frac{m^{N_f}\Lambda^{b_0}}{S^{\hat{N}}} \right)=S \log \left(\frac{\widetilde{\Lambda}^{3\hat{N}}}{S^{\hat{N}}} \right) ,
\end{eqnarray}
where we use the matching condition $\widetilde{\Lambda}^{3\hat{N}}=m^{N_f}\Lambda^{2\hat{N}-N_f}$.
Taking into account for $W(-m_a)$ term discussed in subsection \ref{sub3.2}, we find that (\ref{Aruresult}) exactly concurs with the result $(11)$ in \cite{Sweden} that was obtained by the matrix model. This agreement suggest that the effective superpotential coming from the flavor effect, $W_{\mathrm{eff}}^{flavor}$, corresponds to matrix model free energy with one boundary. 
We can integrate out a massive glueball superfield $S$, using the equation of motion $\partial_S W_{\mathrm{eff}}=0$. This leads to a simple equation,  
\begin{eqnarray}
\partial_s W_{\mathrm{eff}}=-\hat{N}\log \frac{S}{\widetilde{\Lambda}^3}+N_f \log \left[\frac{1}{2}\left(1+\sqrt{1-\frac{4S}{m^2}} \right) \right]=0,
\end{eqnarray}
which becomes
\begin{eqnarray}
\left(\frac{S}{\widetilde{\Lambda}^3} \right)^{2k}-m \left(\frac{S}{\widetilde{\Lambda}^3} \right)^k+\frac{S}{\widetilde{\Lambda}^3}=0 ,\label{fo11}
\end{eqnarray}
where $k\equiv \hat{N}/N_f$. Using (\ref{fo11}) the effective superpotential can be written as
\begin{eqnarray}
W_{\mathrm{eff}}=C \left[(k-1)S+\widetilde{\Lambda}^{3k}S^{-k+1}-\frac{1}{2m^2}\widetilde{\Lambda}^{6k}S^{-2k+2} \right], \label{ADSeff1} 
\end{eqnarray} 
where we write over all constant factor as $C$ for simplicity.  We should use a matching condition $\widetilde{\Lambda}^{3k}=m {\Lambda^{\prime}}^{3k-{1}}$ to recover a matter field. From this relation, (\ref{ADSeff1}) becomes
\begin{eqnarray}
W_{\mathrm{eff}}=C (k-1)S+Cm {\Lambda^{\prime}}^{3k-{1}}S^{-k+1}-\frac{C}{2}{\Lambda^{\prime}}^{6k-{2}} S^{-2k+2}.
\end{eqnarray}
The vacuum expectation value of the field $X=Q\tilde{Q}$ is obtained as a derivative with respect to mass $m$,
\begin{eqnarray} 
X=C{\Lambda^{\prime}}^{3k-{1}}S^{-k+1}.
\end{eqnarray}
We can express the effective superpotential using this new variable $X$: 
\begin{eqnarray}
W_{\mathrm{eff}}=A(N,N_f)\left(\frac{{\Lambda^{\prime}}^{3\hat{N}-N_f}}{\det_{N_f}X} \right)^{\frac{1}{\hat{N}-N_f}}, \label{fo12}
\end{eqnarray}
which is nothing but the Affleck-Dine-Seiberg potential \cite{ADS}. Here in (\ref{fo12}) we ignore tree level superpotential term. This result becomes a more familiar form in the case of $Sp(2N)$ gauge theory with $2N_f$ flavors:
\begin{eqnarray}
W_{\mathrm{eff}}=A(N,N_f)\left(\frac{{\Lambda^{\prime}}^{3(N+1)-N_f}}{\mathrm{Pf}_{N_f}X} \right)^{\frac{1}{N+1-N_f}}.
\end{eqnarray}

Note that this geometric analysis of $U(N)$ gauge theory does not distinguish between $N_f<N$ and $N_f>N$, whereas the gauge theory physics changes drastically. This situation is similar to the matrix model analysis \cite{Sweden,BR,McGreevy}.  Curiously not only $N_f \le N-1$ case but also $N_f\ge N+1$, we obtain the Affleck-Dine-Seiberg superpotential (\ref{fo12}). 

This subject was also discussed in \cite{BR}. The gauge theory with $N_f \ge N+2$ is strongly coupled and the correct description is given by its Seiberg dual.  The superpotential for the dual theory is described in terms of the fields which are dual to the electric meson fields. If we include the dual picture in the computation of effective superpotential, it might resolve this problem.

\section{Seiberg-Witten Curve from Flux}

In this section we reproduce the Seiberg-Witten curve with $N_f$ 
massive flavors from geometry with fluxes.  
Derivation for $U(N)$ gauge theory with no flavor was discussed in 
\cite{CV}. Turning off the fluxes we obtain $\mathcal{N}=2$ quantities such as the Seiberg-Witten curves, coupling constant and BPS mass spectrum. Turning off the fluxes naively means that the tree level superpotential, which break softly $\mathcal{N}=2$ supersymmetry to $\mathcal{N}=1$, goes to zero. Therefore, $\mathcal{N}=2$ supersymmetry is recovered. However, to obtain $\mathcal{N}=2$ information correctly we must consider the order of tree level superpotential and classical value of adjoint chiral multiplet $\Phi$ \cite{CV}.
For $U(N)$ gauge theory, the tree level superpotential is given by
\begin{equation}
W_{\mathrm{tree}}=\sum_{k=1}^{N+1}\frac{g_k}{k}\mathrm{Tr} \Phi^k.
\end{equation}
We assume the vacuum in which gauge group $U(N)$ breaks into $U(1)^N$. That is, $N_i=1$ for $i=1,\ldots , N$ in (\ref{sas2}).
Quantities that do not vanish in the limit $g_{N+1}\to 0$ are the $\mathcal{N}=2$ quantities. In this special case, the massless monopole constraint (\ref{fieldnewmono}) can be written as
\begin{equation}
P^2_N(x)-4\Lambda^{2N-N_f}\prod_{a=1}^{N_f}(x+m_a)=\frac{1}{g_{N+1}}\left({W^{\prime}_{\mathrm{tree}}(x)}^2+f_{N-1}(x) \right).
\end{equation}
This equation means that there is not a massless monopole. Then, on this vacuum, we can express the dual Calabi-Yau geometry (\ref{deformedCY}) as
\begin{equation}
g_{N+1} \biggl( P^2_N(x)-4\Lambda^{2N-N_f}\prod_{a=1}^{N_f}(x+m_a)\biggr)+y^2+z^2+w^2=0.
\end{equation}
After reduction to the $x$-plane and absorbing the $g_{N+1}$ in $y$, we recover the Seiberg-Witten curve
\begin{eqnarray}
y^2=P^2_N(x)-4\Lambda^{2N-N_f}\prod_{a=1}^{N_f}(x+m_a).
\end{eqnarray}

We can discuss the $SO/Sp$ case in a similar way to the $U(N)$ case. 
We consider the tree level superpotential as
\begin{equation}
W_{SO/Sp}=\sum_{k=1}^{N+1}\frac{g_{2k}}{2k}\mathrm{Tr} \Phi^{2k},
\end{equation}
and choose a special vacuum with the breaking pattern $SO(2N)/SO(2N+1)/Sp(2N) \to U(1)^N$. In this special case, the massless monopole constraint equations (\ref{somono}) and (\ref{spmono}) are described as:
\begin{eqnarray}
P_{2N}^2(x)-4\Lambda^{4N-2-2N_f}x^2\prod_{a=1}^{N_f}(x^2-m_a^2)\!\!\!&=&\!\!\! \frac{1}{g_{2N+2}}\left({W^{\prime}_{SO/Sp}(x)}^2+f_{2N} \right) \ \mathrm{for}\ \ SO(2N+1) \\
P_{2N}^2(x)-4\Lambda^{4N-4-2N_f}x^4\prod_{a=1}^{N_f}(x^2-m_a^2)\!\!\! &=&\!\!\! \frac{1}{g_{2N+2}}\left({W^{\prime}_{SO/Sp}(x)}^2+f_{2N} \right) \ \ \mathrm{for}\ \ SO(2N).
\end{eqnarray}
For the $Sp(2N)$ case, we have 
\begin{eqnarray}
\biggl(x^2 P_{2N}+2\Lambda^{2N+2-N_f}\prod_{a=1}^{N_f}m_a \biggr)^2\!\!-4 \Lambda^{2(2N+2-N_f)} \prod_{a=1}^{N_f}(x^2-m_a^2)=\frac{1}{g_{2N+2}}\left({W^{\prime}_{SO/Sp}(x)}^2+f_{2N} \right).
\end{eqnarray}
Then, after the reduction to $x$-plane and absorbing the $g_{2N+2}$, we obtain the Seiberg-Witten curve for 
$SO/Sp$ gauge theory with massive flavors (\ref{SWcurveSO}) and (\ref{SWcurveSp}).

\vspace{1cm}
  
\noindent
{\large{\bf Acknowledgements}}

The author is obliged to Yoshiyuki Watabiki, Hiroyuki Fuji and Takashi Yokono for their stimulating discussion. It is also a pleasure to thank Katsushi Ito for helpful suggestions. The author thanks Minoru Eto and Toshiyuki Hamasaki for their careful reading of this paper and is especially grateful to the referee for helpful comments.


\begin{thebibliography}{100}
 \bibitem{DV}
  R. Dijkgraaf and C. Vafa, ``Matrix models, topological strings, and supersymmetric gauge theories,'' arXiv:hep-th/0206255.
  R. Dijkgraaf and C. Vafa, ``On geometry and matrix models,'' arXiv:hep-th/0207106.
  R. Dijkgraaf and C. Vafa, ``A perturbative window into non-perturbative physics,'' arXiv:hep-th/0208048.

 \bibitem{Sweden}
 R. Argurio, V. L. Campos, G. Ferretti, and R. Heise, ``Exact superpotentials for theories with flavors via a matrix integral,''
arXiv:hep-th/0210291.
 \bibitem{McGreevy}
 J. McGreevy, ``Adding flavor to Dijkgraaf-Vafa,''
arXiv:hep-th/0211009.
 \bibitem{Suzuki}
 H. Suzuki, ``Perturbative Derivation of Exact Superpotential for Meson Fields from Matrix Theories with One Flavor,''
arXiv:hep-th/0211052.
 \bibitem{BR}
 I. Bena and R. Roiban, ``Exact superpotentials in N=1 theories with flavor and their matrix model formulation,''
arXiv:hep-th/0211075.
 \bibitem{NSW}
  S. G. Naculich, H. J. Schnitzer and N. Wyllard, ``Matrix model approach to the N=2 U(N) gauge theory with matter in the fundamental representation
,'' arXiv:hep-th/0211125.

\bibitem{Vafa}
  C. Vafa, ``Superstrings and topological strings at large N,'' J. Math. Phys. \textbf{42} (2001) 2798.
\bibitem{CIV}
  F. Cachazo, K. Intriligator and C. Vafa, ``A large N duality via geometric transition,'' Nucl. Phys. \textbf{B603} (2001) 3.
  \bibitem{CKV}
  F. Cachazo, S. Katz and C. Vafa, ``Geometric transitions and N=1 quiver theories,'' arXiv:hep-th/0108120.
  \bibitem{CFIKV}
  F. Cachazo, B. Fiol, K. Intriligator, S. Katz, and C. Vafa, ``A geometric unification of dualities,'' Nucl.Phys. \textbf{B628} (2002) 3.
\bibitem{EOT}
  J. D. Edelstein, K. Oh and R. Tatar, ``Orientifold, geometric transition
	and large N duality for SO/Sp gauge theories,''
	JHEP.\textbf{0105} (2001) 9.
\bibitem{FO}
  H. Fuji and Y. Ookouchi, ``Confining phase superpotentials for SO/Sp gauge theories via geometric transition,'' arXiv:hep-th/0205301.
  \bibitem{Oh-Tatar}
  K. Dasgupta, K. Oh and R. Tatar
   ``Geometric Transition, Large N Dualities and MQCD Dynamics,''
     Nucl.Phys. {\bf B610} (2001) 331-346, arXiv:hep-th/0105066.
   ``Open/Closed String Dualities 
     and Seiberg Duality from Geometric Transitions in M-theory,''
     arXiv:hep-th/0106040.
  K. Dasgupta, K. Oh, J. Park, and R. Tatar,
   ``Geometric Transition versus Cascading Solution,''
    JHEP {\bf 0201} (2002) 031, arXiv:hep-th/0110050.
  K. Oh and R. Tatar,
   ``Duality and Confinement in N=1 Supersymmetric Theories 
     from Geometric Transitions,''
     arXiv:hep-th/0112040.

  \bibitem{Gukov}
  S. Gukov, C. Vafa and E. Witten, ``CFT's From Calabi-Yau Four-folds,'' Nucl. Phys. B584 (2000) 69, 
 S. Gukov, ``Solitons, Superpotentials and Calibrations,'' Nucl. Phys. B574 (2000) 169.
\bibitem{TV}
  T. R. Taylor and C. Vafa, ``RR flux on Calabi-Yau and partial supersymmetry breaking,'' Phys.Lett. \textbf{474} (2000) 130.
  \bibitem{CV}
  F. Cachazo and C. Vafa, ``N=1 and N=2 geometry from fluxes,'' arXiv: hep-th/0206017.
  \bibitem{AV}
  M. Aganagic and C. Vafa, ``Mirror Symmetry, D-Branes and Counting Holomorphic Discs,'' arXiv:hep-th/0012041.
   \bibitem{Tera9101112}
  A. Hanany and Y. Oz, Nucl. Phys. \textbf{B452} (1995) 238.
  P. C. Argyres, M. R. Plesser and A. D. Shapere, Phys. Rev. Lett. \textbf{75} (1995) 1699.
  P. C. Argyres and A. D. Shapere, Nucl. Phys. \textbf{B461} 437.
  A. Hanany, Nucl. Phys. \textbf{B466} (1996) 85. 
  \bibitem{peskin}
  M. E. Peskin, ``Duality in Supersymmetric Yang-Mills Theory,'' arXiv:hep-th/ 9702094
  \bibitem{ADS}
  I. Affleck, M. Dine and N. Seiberg , Nucl. Phys. \textbf{B241}, 493 (1984).
  \bibitem{DO}
  J. de Boer and Y. Oz, ``Monopole Condensation and Confining Phase of N=1 Gauge Theories Via M Theory Fivebrane,'' Nucl. Phys. \textbf{B511} (1998) 155.
  \bibitem{Ahn}
  C. Ahn, 
      ``Confining Phase of $N=1\ Sp(N_c)$ Gauge Theories 
        via M Theory Fivebrane,'' 
     Phys. Lett. \textbf{B426} (1998) 306.
     , C. Ahn, K. Oh and R. Tatar 
     ``M Theory Fivebrane and Confining Phase of 
       $N=1\ SO(N_c)$ Gauge Theories,'' 
     J. Geom. Phys. \textbf{28} (1998) 163.
  \bibitem{TY}
   S. Terashima 
    ``Supersymmetric gauge theories with classical groups 
      via M theory five-brane,'' 
    Nucl. Phys. \textbf{B526} (1998) 163.

\bibitem{ACHKR}
   S. K. Ashok, R. Corrado, N. Halmagyi, K. D. Kennayay, and C. Romelsberger,  
    ``Unoriented strings, loop equations and N=1 superpotentials from matrix models,''  hep-th/ 0211291.

\bibitem{SV}
S. Sinha and C. Vafa,
``SO and Sp Chern-Simons at large N,''
arXiv:hep-th/0012136.


  
  

\end{thebibliography}
\end{document}